\documentclass[runningheads]{llncs}
\usepackage{graphicx}
\usepackage{comment}
\usepackage{amsmath,amssymb} 
\usepackage{color}
\usepackage{float,multirow,makecell,subfigure,caption,setspace}
\usepackage{booktabs}
\usepackage{amsfonts}
\usepackage{gensymb,textcomp}
\usepackage{bm}
\usepackage{float,multirow,makecell,subfigure,caption,setspace}

\begin{document}
\captionsetup[figure]{labelfont={bf},labelformat={default},labelsep=period,name={Fig.}}
\captionsetup[table]{labelfont={bf},labelformat={default},labelsep=period,name={Table}}
\pagestyle{headings}
\mainmatter
\def\ECCVSubNumber{2160}  

\title{Multi-level Wavelet-based Generative Adversarial Network for Perceptual Quality Enhancement of Compressed Video} 

\author{Jianyi Wang\inst{1}\orcidID{0000-0001-7025-3626} \and
Xin Deng\inst{2}\orcidID{0000-0002-4708-6572} \and
Mai Xu\inst{1,4}\thanks{\scriptsize Corresponding author: Mai Xu}\orcidID{0000-0002-0277-3301} \and
Congyong Chen\inst{3}\orcidID{0000-0003-1682-7435} \and
Yuhang Song\inst{5}\orcidID{0000-0002-7999-0291}}
\titlerunning{MW-GAN}
\authorrunning{J. Wang et al.}
%
\institute{\{School of Electronic and Information Engineering, \and
School of Cyber Science and Technology, \and
College of Software\} Beihang University, Beijing, China  \and
Hangzhou Innovation Institute, Beihang University, Zhejiang, China \and
Department of Computer Science, University of Oxford, United Kingdom\\
\email{\{iceclearwjy,cindydeng,maixu,ndsffx304ccy\}@buaa.edu.cn, yuhang.song@some.ox.ac.uk}}
\maketitle

\begin{abstract}
The past few years have witnessed fast development in video quality enhancement via deep learning.
Existing methods mainly focus on enhancing the objective quality of compressed video while ignoring its perceptual quality.
In this paper, we focus on enhancing the perceptual quality of compressed video.
Our main observation is that enhancing the perceptual quality mostly relies on recovering high-frequency sub-bands in wavelet domain.
Accordingly, we propose a novel generative adversarial network (GAN) based on multi-level wavelet packet transform (WPT) to enhance the perceptual quality of compressed video, which is called multi-level wavelet-based GAN (MW-GAN).
In MW-GAN, we first apply motion compensation with a pyramid architecture to obtain temporal information.
Then, we propose a wavelet reconstruction network with wavelet-dense residual blocks (WDRB) to recover the high-frequency details.
In addition, the adversarial loss of MW-GAN is added via WPT to further encourage high-frequency details recovery for video frames.
Experimental results demonstrate the superiority of our method.
\end{abstract}

\section{Introduction}

\begin{figure}[htbp]
    \begin{center}
        \centerline{\includegraphics[width=1.0\columnwidth]{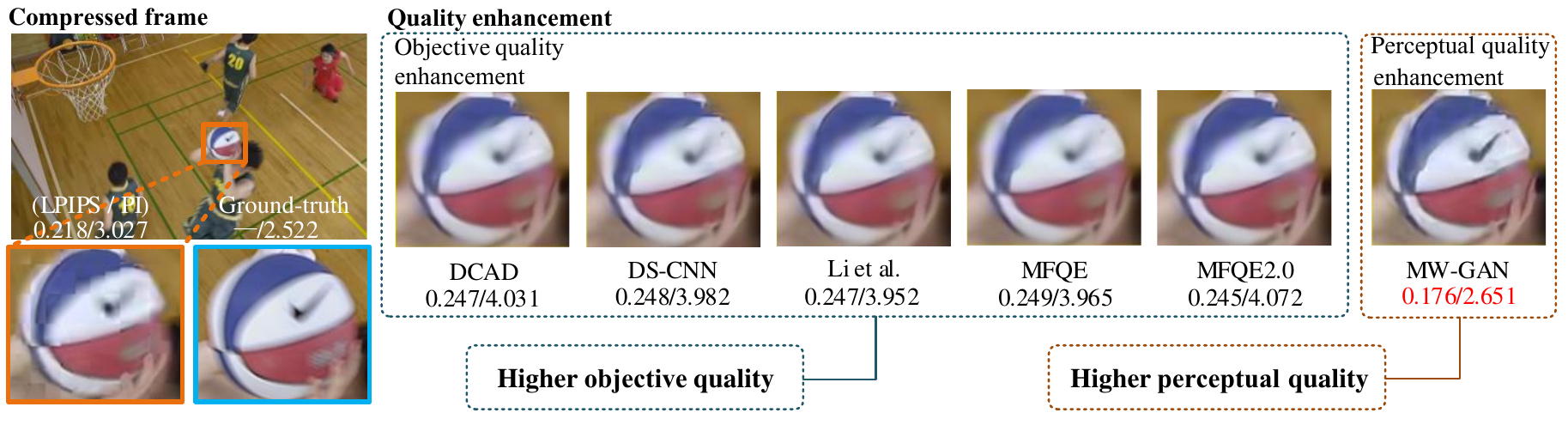}}
        \caption{Objective quality enhancement (traditional works) vs. perceptual quality enhancement(our work). State-of-the-art image \cite{li2017efficient} and video \cite{yang2017decoder,wang2017novel,yang2018multi,guan2019mfqe} quality enhancement methods mainly focus on objective quality enhancement, but ignore perceptual quality, which leads to low perceptual quality with high LPIPS \cite{zhang2018unreasonable} and perceptual index (PI) \cite{blau20182018}. As the first attempt to enhance the perceptual quality of compressed video, our method achieves better perceptual quality with lower LPIPS and PI. Zoom in for best view.}
        \label{figure_1}
    \end{center}
\end{figure}

Nowadays, a large amount of videos are available on the Internet, such as YouTube and Facebook, which exerts huge pressure on the communication bandwidth.
According to the Cisco Visual Networking Index \cite{cvni2019report}, video causes 75\% of Internet traffic in 2017, and this figure is predicted to reach 82\% by 2022.
As a result, video compression has to be applied to save the communication bandwidth \cite{sullivan2012overview}.
However, compressed video inevitably suffers from compression artifacts, which severely degrades the quality of experience (QoE) \cite{li2015weight,bampis2017study}.
Therefore, it is necessary to study on quality enhancement on compressed videos.

Recently, there have been extensive works for enhancing the quality of compressed images and videos \cite{liew2004blocking,foi2007pointwise,jancsary2012loss,dong2015compression,guo2016building,wang2016d3,zhang2017beyond,li2017efficient,tai2017memnet,yang2017decoder,yang2018enhancing,yang2018multi,meng2018mganet}.
Among them, a four-layer convolutional neural network (CNN) called AR-CNN was proposed in \cite{dong2015compression} to improve the quality of JPEG compressed images.
Then, Zhang \textit{et al.} proposed a denoising CNN (DnCNN) \cite{zhang2017beyond} for image denoising and JPEG image deblocking, through a residual learning strategy.
Later, a multi-frame quality enhancement network (MFQE) was proposed in \cite{yang2018multi,guan2019mfqe} to improve the objective quality of compressed video, through leveraging the information from neighboring frames.
Most recently, Yang \textit{et al.} \cite{yang2019quality} have proposed a quality-gated convolutional long short-term memory (QG-ConvLSTM) network, which takes advantage of the bi-directional recurrent structure to fully exploit the useful information in a large range of frames.
Unfortunately, the existing methods mainly focus on improving the objective quality of compressed images/videos while ignoring the perceptual quality.
Actually, high objective quality, i.e., peak signal-to-noise ratio (PSNR), is not always consistent with the human visual system (HVS) \cite{ledig2017photo}.
Besides, according to the perception-distortion tradeoff \cite{blau2018perception}, improving objective quality will inevitably lead to a decrease of perceptual quality.
As illustrated in Figure \ref{figure_1}, although the frames generated by state-of-the-art methods \cite{li2017efficient,yang2017decoder,wang2017novel,yang2018multi,guan2019mfqe} have high PSNR values, they are not perceptually photorealistic with high LPIPS value \cite{zhang2018unreasonable} and perceptual index (PI) \cite{blau20182018}, due to the lack of high-frequency details and fine textures.

In this paper, we propose a multi-level wavelet-based generative adversarial network (MW-GAN) for perceptual quality enhancement of compressed video, which recovers the high-frequency details via wavelet packet transform (WPT) \cite{mallat1999wavelet} at multiple levels.
The key insight to adopt WPT is that the high-frequency details are usually missing due to the compression and they can be regarded as the high-frequency sub-bands after WPT. Our MW-GAN has a generator and a discriminator.
Specifically, the generator is composed of two main components: motion compensation and wavelet reconstruction.
Motion compensation is first developed to compensate the motion between the target frame and its neighbors with a pyramid architecture following \cite{guan2019mfqe}.
Then, the wavelet reconstruction network consisting of a bunch of wavelet-dense residual blocks (WDRB) is adopted to reconstruct the sub-bands of the target frame.
Besides, WPT is also considered in the discriminator of MW-GAN, such that the generator is further encouraged to recover high-frequency details.
As shown in Figure \ref{figure_1}, our method is able to generate photorealistic frames with sufficient texture details.

To the best of our knowledge, our work is the first attempt to enhance the perceptual quality of compressed video, using a wavelet-based GAN.
The main contributions of this paper are as follows:
(1) We investigate that the high-frequency sub-bands in wavelet domain is highly related to the perceptual quality of compressed video.
(2) We propose a novel network architecture called MW-GAN, which learns to recover the high-frequency information in wavelet domain for perceptual quality enhancement of compressed video.
(3) Extensive experiments have been conducted to demonstrate the ability of the proposed method in enhancing the perceptual quality of compressed video.

\section{Related Work}
\textbf{Quality enhancement of compressed images/videos.}
In the past few years, extensive works \cite{liew2004blocking,foi2007pointwise,jancsary2012loss,jung2012image,chang2013reducing,dong2015compression,guo2016building,wang2016d3,zhang2017beyond,li2017efficient,cavigelli2017cas,tai2017memnet,yang2017decoder,yang2018multi,guan2019mfqe} have been proposed to enhance the objective quality of compressed images/videos.
Specifically, for compressed images, Foi \textit{et al.} \cite{foi2007pointwise} applied point-wise shape-adaptiveDCT (SADCT) to reduce the blocking and ringing effects caused by JPEG compression.
Then, regression tree fields (RTF) were adopted in \cite{jancsary2012loss} to reduce JPEG image blocking effects.
Moreover, \cite{jung2012image} and \cite{chang2013reducing} utilized sparse coding to remove JPEG artifacts.
Recently, deep learning has also been successfully applied for quality enhancement.
Particularly, Dong \textit{et al.} \cite{dong2015compression} proposed a four-layer AR-CNN to reduce the JPEG artifacts of images.
Afterwards, $\textbf{D}^3$ \cite{wang2016d3} and deep dual-domain convolutional network (DDCN) \cite{guo2016building} were proposed for enhancing the quality of JPEG compressed images, utilizing the prior knowledge of JPEG compression.
Later, DnCNN was proposed in \cite{zhang2017beyond} for multiple tasks of image restoration, including quality enhancement.
Li \textit{et al.} \cite{li2017efficient} proposed a 20-layer CNN for enhancing image quality, which achieves state-of-the-art performance on objective quality enhancement of compressed images.

For the compressed videos, Wang \textit{et al.} \cite{wang2017novel} proposed a deep CNN-based auto decoder (DCAD) which applies 10 CNN layers to reduce the distortion of compressed video.
Later, DS-CNN \cite{yang2017decoder} was proposed for video quality enhancement, in which two sub-networks are used to reduce the artifacts of intra- and inter-coding frames, respectively.
Besides, Yang \textit{et al.} \cite{yang2018multi} proposed a multi-frame quality enhancement network (MFQE) to take advantage of neighbor high-quality frames.
Most recently, a quality-gated convolutional long short-term memory (QG-ConvLSTM) \cite{yang2019quality} network was proposed to enhance video quality via learning the ``forget'' and ``input'' gates in the ConvLSTM \cite{xingjian2015convolutional} cell from quality-related features.
Then, Guan \textit{et al.} \cite{guan2019mfqe} proposed a new method for multi-frame quality enhancement called MFQE 2.0, which is an extended version of \cite{yang2018multi} with substantial improvement and achieved state-of-the-art performance.
All the above methods aim at minimizing the pixel-wise loss, such as mean square error (MSE), to obtain high objective quality.
However, according to \cite{ledig2017photo}, MSE cannot always reflect the perceptually relevant differences.
Actually, minimizing MSE encourages finding pixel-wise averages of plausible solutions, leading to overly-smooth images/videos with poor perceptual quality \cite{mathieu2015deep,johnson2016perceptual,dosovitskiy2016generating}.

\textbf{Perception-driven super-resolution.}
To the best of our knowledge, there exists no work on perceptual quality enhancement of compressed video.
The closest work to ours is the perception-driven image/video super-resolution that aims to restore perception-friendly high-resolution images/videos from low-resolution ones.
Specifically, Johnson \textit{et al.} \cite{johnson2016perceptual} proposed to minimize perceptual loss defined in the feature space to enhance the perceptual image quality in single image super-resolution.
Then, contextual loss \cite{mechrez2018maintaining} was developed to generate images with natural image statistics via using an objective that focuses on the feature distribution.
SRGAN \cite{ledig2017photo} was proposed to generate natural images using perceptual loss and adversarial loss.
Sajjadi \textit{et al.} \cite{sajjadi2017enhancenet} developed a similar method with the local texture matching loss.
Later, Wang \textit{et al.} \cite{wang2018recovering} proposed spatial feature transform to effectively incorporate semantic prior in an image and improve the recovered textures.
They also developed an enhanced version of SRGAN (ESRGAN) \cite{wang2018esrgan} with a deeper and more efficient network architecture.
Most recently, TecoGAN \cite{chu2018temporally} was proposed to achieve state-of-the-art performance in video super-resolution, in which a spatial-temporal discriminator and a ping-pong loss are applied.
Similar to super-resolution, perceptual quality is also important for quality enhancement of compressed video.
To the best of our knowledge, our MW-GAN in this paper is the first attempt in this direction.

\begin{figure}[htbp]
    \begin{center}
        \centerline{\includegraphics[width=1.0\columnwidth]{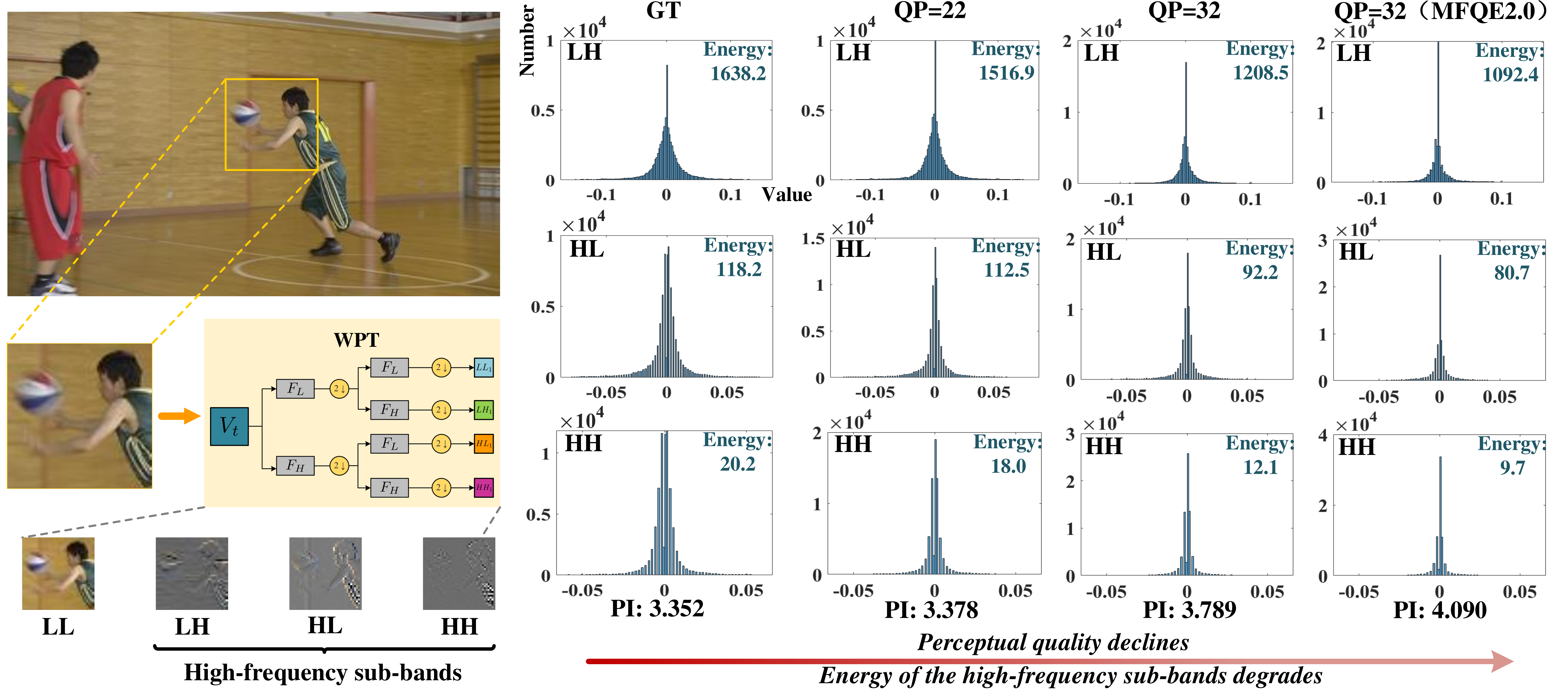}}
        \caption{Histograms of the wavelet sub-bands (zoom in for details). Compressed frames with higher QP have lower perceptual quality (i.e., higher PI) while high-frequency wavelets fade with degraded energy alongside the increase of the QP value. Besides, the state-of-the-art method \cite{guan2019mfqe} obtains low perceptual quality, due to the failure of enhancing high-frequency sub-bands. Similar results can be found for other state-of-the-art methods \cite{yang2018multi,li2017efficient,yang2017decoder,wang2017novel}.}
        \label{wpt_exp}
    \end{center}
\end{figure}

\section{Motivation for WPT}
The wavelet transform allows the multi-resolution analysis of images \cite{jawerth1994overview}, and it can decompose an image into multiple sub-bands, i.e., low- and high-frequency sub-bands. As verified in \cite{deng2019wavelet}, the high-frequency sub-bands play an important role in enhancing the perceptual quality of a super-resolved image. Here,
we further verify that the high-frequency sub-bands are also crucial for the perceptual quality enhancement of compressed video.
Specifically, given a lossless video frame, a series of compressed frames are obtained via HM16.5 \cite{sullivan2012overview} at low-delay configuration, with the quantization parameter (QP) of $22$, $27$, $32$ and $37$.
The corresponding wavelet sub-bands are achieved via WPT using \textit{haar} filter at one level.
For each frame, we obtain four sub-bands called LL, LH, HL and HH.
Here, LL is the low-frequency sub-band, LH, HL and HH are the sub-bands with high-frequency information at horizontal, vertical and diagonal directions, respectively.
Figure \ref{wpt_exp} shows the histograms of wavelet coefficients of the LH, HL and HH sub-bands. As we can see, the compressed frames with higher QP values tend to have lower perceptual quality (i.e., higher PI) and insufficient high-frequency wavelet coefficients. Moreover, although the state-of-the-art method \cite{guan2019mfqe} can obtain high objective quality, it fails to recover the high-frequency sub-bands in wavelet domain, leading to low perceptual quality.
In order to measure the richness of high-frequency wavelet coefficients, we further calculate the wavelet energy by summing up their squared coefficients.
As shown in Table \ref{subband_result},  the compressed frames with higher QP values tend to have lower wavelet energy in high-frequency sub-bands.
All these demonstrate that the high-frequency wavelet sub-bands play an important role in enhancing the perceptual quality of compressed video.

\begin{table}
    \newcommand{\tabincell}[2]{\begin{tabular}{@{}#1@{}}#2\end{tabular}}
    \begin{center}
        \caption{Wavelet energy of the high-frequency sub-bands (LH, HL, HH) per frame across the MFQE2.0 dataset \cite{guan2019mfqe}. Frames with higher QP values tend to have lower wavelet energy in high-frequency sub-bands.}
        \label{subband_result}
        \resizebox{0.9\textwidth}{!}{
        \begin{tabular}{c|cccc|cccc|c}
            \hline  Wavelet & \multicolumn{4}{c|}{Compressed frame} & \multicolumn{4}{c|}{MFQE2.0 \cite{guan2019mfqe}} & Ground-\\
            \cline{2-9} sub-bands & QP=$22$ & QP=$27$ & QP=$32$ & QP=$37$ & QP=$22$ & QP=$27$ & QP=$32$ & QP=$37$ & truth \\
            \hline  LH & 1106.8 & 1043.3 & 956.1 & 835.4 & 872.0 & 844.2 & 809.9 & 715.9 & 1189.7 \\
            \hline  HL & 1189.3 & 1130.2 & 1048.1 & 940.5 & 994.3 & 973.9 & 937.8 & 872.1 & 1273.0 \\
            \hline  HH & 138.4 & 115.9 & 93.8 & 70.7 & 130.9 & 110.4 & 92.6 & 65.3 & 176.2 \\
            \hline
        \end{tabular}
        }
    \end{center}
\end{table}

Based on the above investigation, we propose our MW-GAN as follows.
(1) Since high-frequency wavelet sub-bands are crucial for the perceptual quality, the generator of our MW-GAN directly outputs these wavelet sub-bands.
(2) To restore the high-frequency wavelet sub-bands as much as possible, a wavelet-dense residual block (WDRB) is proposed in our MW-GAN to further recover the wavelet sub-bands.
(3) The WPT is also applied in the discriminator of our MW-GAN for perceptual quality enhancement by encouraging the generated results to be indistinguishable from the ground-truth in wavelet domain.

\section{The Proposed MW-GAN}

\begin{figure}[htbp]
    \begin{center}
        \centerline{\includegraphics[width=1.0\columnwidth]{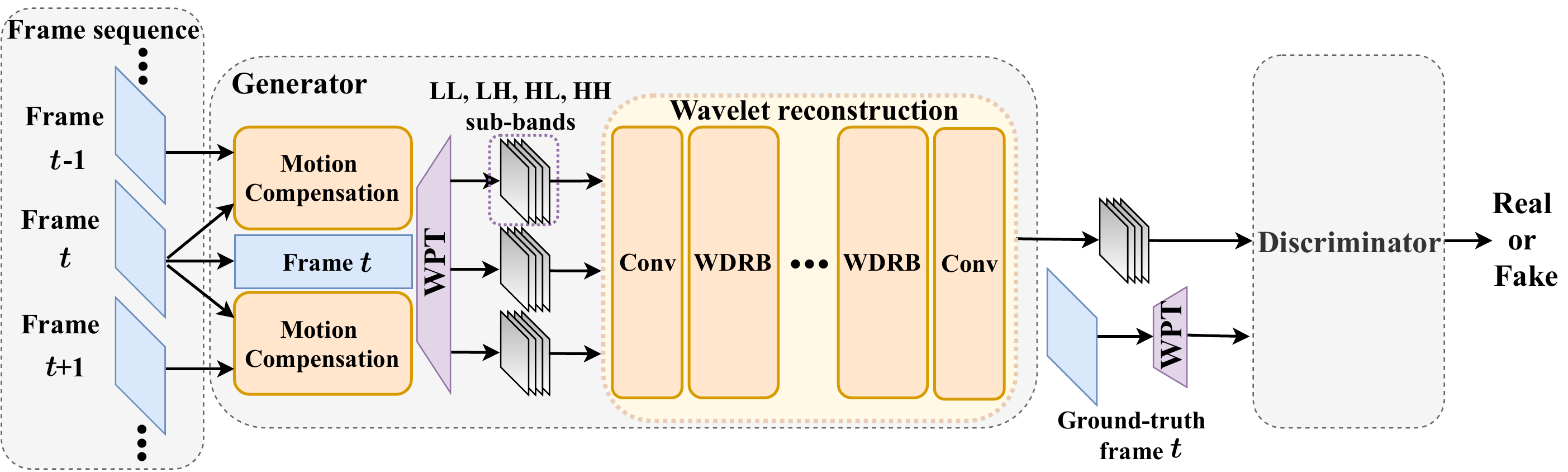}}
        \caption{Framework of our MW-GAN. Perceptual quality is enhanced via recovering wavelet sub-bands in wavelet domain. Specifically, we propose a generator with motion composition and WDRB to capture both temporal and high-frequency information. Then a multi-level wavelet-based discriminator is proposed to evaluate generated results in both pixel and wavelet domain.}
        \label{framework}
    \end{center}
\end{figure}

The architecture of the proposed MW-GAN is shown in Figure \ref{framework}.
To enhance the perceptual quality of a video frame $\mathbf{V}_{t}$, we simultaneously train a generator $G_{\bm{\theta}_G}$ parameterized by $\bm{\theta}_G$ and a discriminator $D_{\bm{\theta}_D}$ parameterized by $\bm{\theta}_D$ in an adversarial manner.
The ultimate goal is to obtain the enhanced frame $\hat{\mathbf{O}}_{t}$ with high perceptual quality, similar to its ground-truth $\mathbf{O}_{t}$.
To take advantage of the temporal information in adjacent frames, the generator $G_{\bm{\theta}_G}$ takes both $\mathbf{V}_{t}$ and its neighbor frames $\{\mathbf{V}_{t \pm n}\}^{N}_{n =1}$ as inputs.
The details about the generator $G_{\bm{\theta}_G}$ are introduced in Section \ref{G_section}.
To further encourage high perceptual quality, we propose a multi-level wavelet-based discriminator, which distinguishes the generated wavelet sub-bands from the ground-truths, as introduced in Section \ref{D_section}.
Finally, we present the loss functions to train the MW-GAN in Section \ref{loss_section}.

\subsection{Multi-level wavelet-based generator}
\label{G_section}
The generator $G_{\bm{\theta}_G}$ of our MW-GAN is mainly composed of two parts: motion compensation and wavelet reconstruction.
Given the $2N+1$ frames as inputs, we first perform motion compensation across frames to align content across frames.
After that, wavelet reconstruction is applied to reconstruct the target wavelet sub-bands using the information of sub-bands from both current and compensated frames.
Finally, the output reconstructed sub-bands are used to obtain the enhanced frame through IWPT.

\textbf{Motion compensation}:
To take advantage of the temporal information across frames, we adopt the motion compensation network with a pyramid architecture following \cite{yang2018multi}.
To obtain better performance, we further make some modifications.
First, for flow estimation, it is crucial to have large-size filters in the first few layers of the network to capture intense motion.
Thus, the kernel sizes of the first two convolutional layers at each pyramid level are $7 \times 7$ and $5 \times 5$, respectively, instead of $3\times 3$ in \cite{yang2018multi}.
Second, we replace the pooling operation at each pyramid level by WPT.
Since WPT is invertible, this pooling operation is able to make all information of the original frames be preserved, meanwhile efficiently enlarging the receptive field for flow estimation.

\begin{figure}
    \begin{center}
        \subfigure{
        \begin{minipage}[t]{0.8\linewidth}
        \centering
        \includegraphics[width=2.8in]{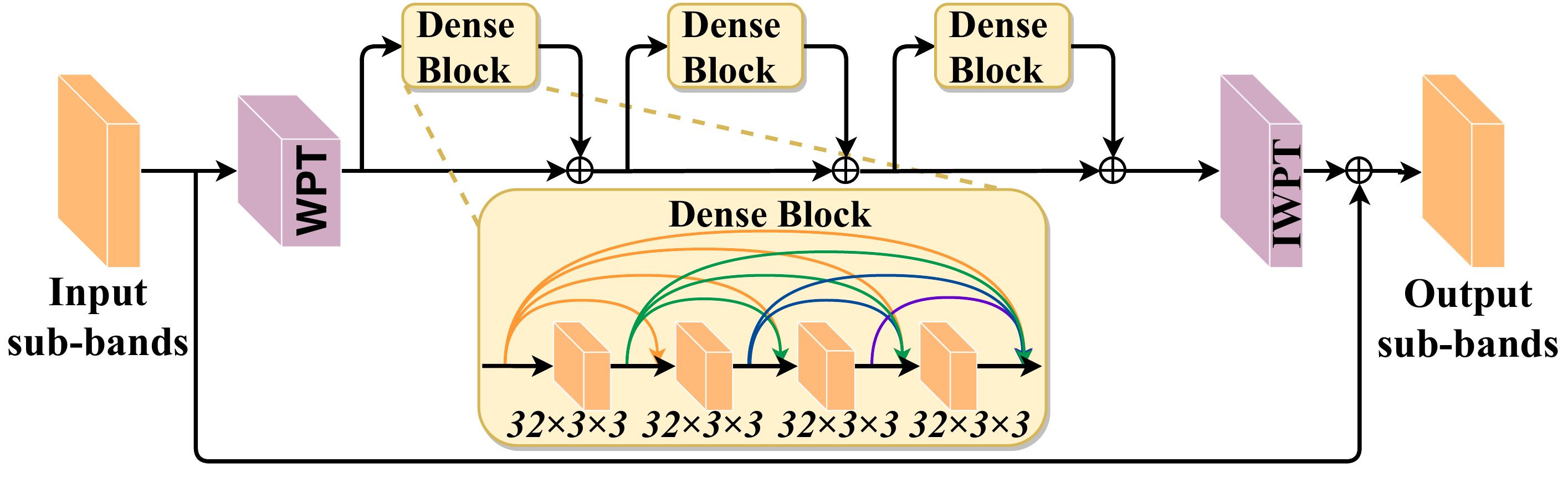}
        \end{minipage}%
        } \\
        \subfigure{
        \begin{minipage}[t]{0.8\linewidth}
        \centering
        \includegraphics[width=2.8in]{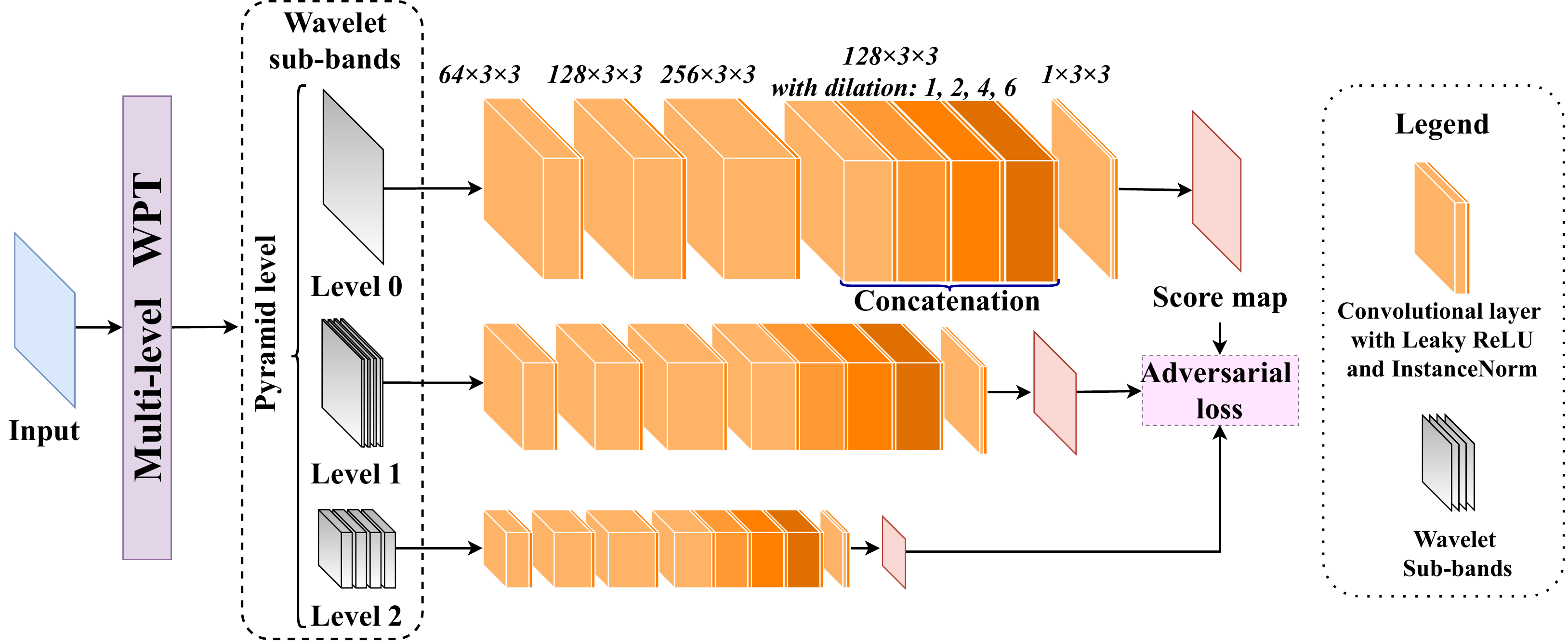}
        \end{minipage}%
        }%
        \end{center}
    \caption{Details about our framework. Top: Architecture of the proposed WDRB; Down: Architecture of the multi-level wavelet-based discriminator in MW-GAN.}
    \label{G_network}
\end{figure}

\textbf{Wavelet reconstruction}:
Given the current and compensated frames, we further propose a wavelet reconstruction network to reconstruct the wavelet sub-bands $\hat{\mathbf{S}}_t$, which contains a cascade of wavelet-dense residual blocks (WDRB) .
Specifically, WPT is first applied to generate sub-bands as inputs.
Then, several convolutional layers are applied to extract corresponding feature maps.
To further capture high-frequency details and reduce the computational cost, we develop the wavelet-based residual block (WDRB) with a residual-in-residual structure, as shown in Figure \ref{G_network}.
Similar to \cite{wang2018esrgan}, in our WDRB, several dense blocks are applied in the main path of the residual block to enhance the feature representation with high capacity.
In addition, we adopt WPT and IWPT in the main path to learn the residual in wavelet domain.
With this simple yet effective design, the receptive field is further enlarged without losing information, thus enabling the network to capture more high-frequency details.
The efficiency of the WDRB is also investigated in the ablation study.
The final output of $G_{\bm{\theta}_G}$ are the wavelet sub-bands $\hat{\mathbf{S}}_t$, and the enhanced frame $\hat{\mathbf{O}}_{t}$ can be obtained by combining $\hat{\mathbf{S}}_t$ via IWPT.
Assuming that $\omega^{-1}(\cdot)$ denotes the IWPT operation, the output of the generator can be summarized as follows,
\begin{equation}
\label{g_output}
\begin{aligned}
& \hat{\mathbf{S}}_t= G_{\bm{\theta}_G}(\mathbf{V}_t, \{\mathbf{V}_{t \pm n}\}^N_{n=1}),\\
& \hat{\mathbf{O}}_{t} = \omega^{-1}(\hat{\mathbf{S}}_t).
\end{aligned}
\end{equation}

\subsection{Multi-level wavelet-based discriminator}
\label{D_section}
In this section, a multi-level wavelet-based discriminator is proposed to encourage the generated results indistinguishable from the ground-truth.
The structure of our discriminator is illustrated in Figure \ref{G_network}.
Specifically, the discriminator takes the generated frames (obtained via IWPT) or the real ones (ground-truth) as input.
Then, WPT is applied to both generated and real frames for obtaining their wavelet sub-bands at multiple levels.
After that, wavelet sub-bands of different levels are fed into the corresponding pyramid level.
Each pyramid level employs a fully convolutional architecture, in which a group of convolutional layers with different dilations are applied to extract features with various receptive fields.
Subsequently, the input wavelet sub-bands at each level are mapped into a single channel of the score map with the same size as the input, which measures the similarity between the generated and real frames.
Let $\omega_l(\cdot)$ denote the $l$-th level WPT.
The output of the discriminator is a set of score maps of different sizes, which can be formulated as follows,

\begin{equation}
\label{d_output}
\begin{split}
D^l_{\bm{\theta}_D}(\mathbf{O}_{t}) = f_l(\omega_l(\mathbf{O}_{t})),
\end{split}
\end{equation}
where $f_l(\cdot)$ is the corresponding map function at the $l$-th level of the discriminator, $L$ is the total number of the pyramid levels and $D^l_{\bm{\theta}_D}(\mathbf{O}_{t})$ is the output score map  at $l$-th level. Note that the $l=0$ indicates that there is no WPT operation.

The advantage of the proposed multi-level wavelet-based discriminator is efficiency and efficacy.
(1) The computational cost can be significantly reduced by adopting WPT, since the calculation of wavelet sub-bands can be regarded as forwarding through a single convolutional layer.
In contrast, the traditional methods need to extract multi-scale features via a sequence of convolutional layers, thus consuming expensive computational complexity.
(2) Our discriminator can distinguish the generated and real frames in both pixel ($l=0$) and wavelet domains ($l>0$), leading to better generative results.

\subsection{Loss functions}
\label{loss_section}
In this section, we propose a loss function for training the MW-GAN.
Specifically, we formulate the loss of the generator $L_G(\bm{\theta}_M;\bm{\theta}_G)$ as the weighted sum of a motion loss $L_{M}(\bm{\theta}_M)$, a wavelet loss $L_{W}(\bm{\theta}_G)$ and an adversarial loss $L_{\rm Adv}(\bm{\theta}_G)$. Then, $L_G(\bm{\theta}_M;\bm{\theta}_G)$ can be written as,
\begin{equation}
\label{total_l}
L_G(\bm{\theta}_M;\bm{\theta}_G) = L_{W}(\bm{\theta}_G) + \alpha L_{M}(\bm{\theta}_M) + \beta L_{\rm Adv}(\bm{\theta}_G),
\end{equation}
where $\alpha$ and $\beta$ are the coefficients to balance different loss terms.
In the following, we explain the different components of the loss functions.

\textbf{Motion loss}: Since it is hard to obtain the ground-truth of optical flow, the motion compensation network $M_{\bm{\theta}_M}$ parameterized by $\bm{\theta}_M$ is directly trained under the supervision of the ground-truth $\mathbf{O}_t$.
That is, the neighboring frames $\{\mathbf{V}_{t \pm n}\}^{N}_{n =1}$ are first wrapped using the optical flow estimated by $M_{\bm{\theta}_M}$; then the loss between the compensated frames $\{M_{\bm{\theta}_M}(\mathbf{V}_{t\pm n})\}^{N}_{n = 1}$ and the ground-truth $\mathbf{O}_t$ is minimized.
Here, we adopt Charbonnier penalty function \cite{lai2017deep} as the motion loss, defined by:
\begin{equation}
\label{motion_l}
L_{M}(\bm{\theta}_M) =  \frac{1}{2N}\sum\limits^{N}_{\substack{n=1} }\sqrt{\left\|M_{\bm{\theta}_M}(\mathbf{V}_{t\pm n})-\mathbf{O}_t\right\|^2_F + \epsilon^2_m},
\end{equation}
where $\epsilon_m$ is a scaling parameter and is empirically set to $1 \times 10^{-3}$.

\textbf{Wavelet loss}: The wavelet loss function models how close the predicted sub-bands $\hat{\bm{S}}_t$ are to the ground truth $\bm{S}_t$. It is defined by a weighted Charbonnier penalty function in wavelet domain as follows:
\begin{equation}
\label{wavelet_l}
\begin{aligned}
L_{W}(\bm{\theta}_G)&=\sqrt{\left\|\mathbf{W}^{1/2} \odot (\hat{\mathbf{S}}_t-\mathbf{S}_t) \right\|^2_F + \epsilon^2_w},
\end{aligned}
\end{equation}
where $\odot$ represents dot product, $\hat{\mathbf{S}}_t=\{\hat{\mathbf{s}}^1_t, \hat{\mathbf{s}}^2_t, \ldots , \hat{\mathbf{s}}^{n_w}_t\}$ are the predicted sub-bands with the number of $n_w$ in total and $\mathbf{S}_t$ are their corresponding ground-truths.
In addition, $\mathbf{W}=\{w_1, w_2, \ldots, w_{n_w} \}$ is the weight matrix to balance the importance of each sub-band and $\epsilon_w$ is a scaling parameter set to $1 \times 10^{-3}$.

\textbf{Adversarial loss}:
In addition to the above two loss functions, another important loss of the GAN is the adversarial loss.
Note that we follow \cite{mao2017least} to adopt $\ell_2$ loss for training.
Recall that $D^l_{\bm{\theta}_D}(\mathbf{O}_{t})$ represents the outputs of the discriminator in \eqref{d_output} at the $l$-th level, taking $\mathbf{O}_t$ as input.
Then, the discriminator loss can be defined as:
\begin{equation}
\begin{split}
L_D(\bm{\theta}_D) = \frac{1}{2L}\mathbb{E}_{\mathbf{O}_{t}}[\sum\limits^{L-1}_{l=0}\|D^l_{\bm{\theta}_D}(\mathbf{O}_{t})-\mathbf{1}\|^2_F] +\frac{1}{2L}\mathbb{E}_{\hat{\mathbf{O}}_{t}}[\sum\limits^{L-1}_{l=0}\|D^l_{\bm{\theta}_D}(\hat{\mathbf{O}}_{t})\|^2_F],
\end{split}
\end{equation}
where $\mathbb{E}_{\hat{\mathbf{O}}_t}[\cdot]$ represents the average for all $\hat{O}_t$ in the mini-batch and $L$ is the number of the pyramid levels in the discriminator.
Note that $\mathbf{1} \in \mathcal{R}^{W_l \times H_l}$, where $W_l$ and $H_l$ are the width and height of the score map at the $l$-th level, respectively.
Symmetrically, the adversarial loss for generator is as follow:
\begin{equation}
L_{\rm Adv}(\bm{\theta}_G) = \frac{1}{2L}\mathbb{E}_{\hat{\mathbf{O}}_{t}}[\sum\limits^{L-1}_{l=0}\|D^l_{\bm{\theta}_D}(\hat{\mathbf{O}}_{t}(\bm{\theta}_G))-\mathbf{1}\|^2_F].
\end{equation}

\section{Experiments}
In this section, the experimental results are presented to validate the effectiveness of our MW-GAN method.
Section \ref{exp_settings} introduces the experimental settings.
Section \ref{exp_pi} compares the results between our method and other state-of-the-art methods over the test sequences of JCT-VC \cite{ohm2012comparison}.
In Section \ref{exp_mos}, the mean opinion score (MOS) test is performed to compare the subjective quality of videos by different methods.
Finally, the ablation study is presented in Section \ref{exp_abla}.

\subsection{Settings}
\label{exp_settings}
\textbf{Datasets}:
We train the MW-GAN model using the database introduced in \cite{guan2019mfqe}.
Following \cite{guan2019mfqe}, except the $18$ common test sequences of Joint Collaborative Team on Video Coding (JCT-VC) \cite{ohm2012comparison}, the other $142$ sequences are randomly divided into non-overlapping training set ($106$ sequences) and validation set ($36$ sequences).
All $160$ sequences are compressed by HM16.5 \cite{sullivan2012overview} under Low-Delay configuration, setting QP to $32$.
Note that different from existing methods \cite{li2017efficient,cavigelli2017cas,tai2017memnet,yang2017decoder,yang2018multi,yang2019quality,guan2019mfqe} that train models for each QP individually, we only train our MW-GAN under QP$=32$, and we test on both QP$=32$ and QP$=37$.
The results show the generalization ability of our method.

\textbf{Parameter settings}:
Here, we mainly introduce the settings and hyperparameters of our experiments.
Specifically, the total number of levels $L$ of each component in MW-GAN is set to $3$ and the generator takes 3 consecutive frames as inputs for a better tradeoff between performance and efficiency.
The training process is divided into two stages.
We first train our model without adversarial loss (i.e., $\beta=0$).
The motion loss weight $\alpha$ is initialed as $10$ and decayed by a factor of $10$ every $5 \times 10^4$ iterations to encourage the learning of motion compensation first.
The iteration number and mini-batch size are $3 \times 10^5$ and 32, respectively, and the input frames are cropped to $256 \times 256$.
The Adam algorithm \cite{kingma2014adam} with the step size of $2.5\times10^{-4}$ is adopted; the learning rate is initialized as $2 \times 10^{-4}$ and decayed by a factor of $2$ every $1 \times 10^5$ iterations.
In the second stage, the pre-trained model is employed as an initialization for the generator.
The generator is trained using the loss function in \eqref{total_l} with $\alpha = 1 \times 10^{-2}$ and $\beta = 5 \times 10^{-3}$.
The iteration number and mini-batch size are $6 \times 10^5$ and 32, respectively, and the cropped size is reduced to $128 \times 128$.
The learning rate is set to $1 \times 10^{-4}$ and halved every $1 \times 10^5$ iterations.
Note that the above hyperparameteres are tuned over the training set.
We apply the Adam algorithm \cite{kingma2014adam} and alternately update the generator and discriminator network until convergence.
The wavelet sub-bands are obtained by wavelet packet decomposition with $haar$ filter.
More details can be found in the supplementary material.

\subsection{Quantatative comparison}
\label{exp_pi}

\begin{table}
    \newcommand{\tabincell}[2]{\begin{tabular}{@{}#1@{}}#2\end{tabular}}
    \begin{center}
        \caption{Overall $\Delta \rm {LPIPS}$ and $\Delta \rm {PI}$ between enhanced and compressed frames on the test set of JCT-VC \cite{ohm2012comparison} at QP$=32$ and QP$=37$. Our MW-GAN achieves the best perceptual quality across all the test sequences.}
        \label{metric_table}
\resizebox{1.0\textwidth}{!}{
        \begin{tabular}{|c|c|c|cc|cc|cc|cc|cc|cc|cc|}
            \hline \multirow{2}{*}{QP} & \multicolumn{2}{c|}{\multirow{2}{*}{Video sequence}} & \multicolumn{2}{c}{Li \textit{et al.} \cite{li2017efficient}} & \multicolumn{2}{c}{DCAD \cite{wang2017novel}} & \multicolumn{2}{c}{DS-CNN \cite{yang2017decoder}} & \multicolumn{2}{c}{MFQE \cite{yang2018multi}} & \multicolumn{2}{c}{MFQE 2.0 \cite{guan2019mfqe}} & \multicolumn{2}{c|}{Ours}\\
            \cline{4-15}& \multicolumn{2}{c|}{\multirow{2}{*}{}} & $\Delta \rm{LPIPS}$ & $\Delta \rm{PI}$ & $\Delta \rm{LPIPS}$ & $\Delta \rm{PI}$ & $\Delta \rm{LPIPS}$ & $\Delta \rm{PI}$ & $\Delta \rm{LPIPS}$ & $\Delta \rm{PI}$ & $\Delta \rm{LPIPS}$ & $\Delta \rm{PI}$ & $\Delta \rm{LPIPS}$ & $\Delta \rm{PI}$\\
            \hline \multirow{19}{*}{32}& \multirow{2}{*}{A} & \textit{Traffic} & 0.021 & 0.501 & 0.019 & 0.419 & 0.017 & 0.373 & 0.016 & 0.441 & 0.014 & 0.430 & \textbf{-0.032} & \textbf{-1.123} \\
            \cline{3-15}  & & \textit{PeopleOnStreet} & 0.020 & 0.865 & 0.020 & 0.668 & 0.019 & 0.663 & 0.017 & 0.807 & 0.017 & 0.794 & \textbf{-0.020} & \textbf{-0.558}\\
           \cline{2-15} & \multirow{5}{*}{B} & \textit{Kimono} & 0.038 & 0.479 & 0.034 & 0.403 & 0.033 & 0.332 & 0.034 & 0.440 & 0.036 & 0.443 & \textbf{-0.069} & \textbf{-1.561}\\
            \cline{3-15} & & \textit{ParkScene} & 0.010 & 0.299 & 0.010 & 0.377 & 0.009 & 0.291 & 0.012 & 0.239 & 0.010 & 0.346 & \textbf{-0.032} & \textbf{-0.159}\\
            \cline{3-15} & & \textit{Cactus} & 0.032 & 0.264 & 0.028 & 0.264 & 0.027 & 0.236 & 0.028 & 0.248 & 0.028 & 0.274 & \textbf{-0.109} & \textbf{-0.953}\\
            \cline{3-15} & & \textit{BQTerrace} & 0.028 & 0.384 & 0.025 & 0.413 & 0.025 & 0.362 & 0.026 & 0.418 & 0.026 & 0.447 & \textbf{-0.099} & \textbf{-0.326}\\
            \cline{3-15} & & \textit{BasketballDrive} & 0.036 & 0.534 & 0.031 & 0.501 & 0.032 & 0.462 & 0.032 & 0.595 & 0.032 & 0.618 & \textbf{-0.106} & \textbf{-0.921}\\
            \cline{2-15} & \multirow{4}{*}{C} & \textit{RaceHorses} & 0.023 & 0.504 & 0.022 & 0.536 & 0.021 & 0.469 & 0.025 & 0.489 & 0.027 & 0.566 & \textbf{-0.021} & \textbf{-0.579}\\
            \cline{3-15} &  & \textit{BQMall} & 0.022 & 0.457 & 0.019 & 0.507 & 0.018 & 0.426 & 0.021 & 0.467 & 0.021 & 0.499 & \textbf{-0.033} & \textbf{-0.908}\\
            \cline{3-15} &  & \textit{PartyScene} & 0.032 & 0.436 & 0.027 & 0.381 & 0.024 & 0.347 & 0.025 & 0.408 & 0.025 & 0.412 & \textbf{-0.075} & \textbf{-0.765}\\
            \cline{3-15} &  & \textit{BasketballDrill} & 0.027 & 0.782 & 0.023 & 0.926 & 0.026 & 0.842 & 0.027 & 0.772 & 0.025 & 0.929 & \textbf{-0.047} & \textbf{-0.532}\\
            \cline{2-15} & \multirow{4}{*}{D} & \textit{RaceHorses} & 0.019 & 0.628 & 0.018 & 0.657 & 0.017 & 0.581 & 0.021 & 0.657 & 0.021 & 0.685 & \textbf{-0.005} & \textbf{-0.199}\\
            \cline{3-15} & & \textit{BQSquare} & 0.012 & 0.285 & 0.012 & 0.599 & 0.011 & 0.382 & 0.013 & 0.344 & 0.011 & 0.570 & \textbf{-0.037} & \textbf{-0.525}\\
            \cline{3-15} & & \textit{BlowingBubbles} & 0.013 & 0.492 & 0.014 & 0.585 & 0.011 & 0.458 & 0.012 & 0.394 & 0.009 & 0.472 & \textbf{-0.039} & \textbf{-0.241}\\
            \cline{3-15} & & \textit{BasketballPass} & 0.020 & 0.537 & 0.016 & 0.564 & 0.016 & 0.517 & 0.020 & 0.574 & 0.019 & 0.594 & \textbf{-0.021} & \textbf{-0.794}\\
            \cline{2-15} & \multirow{3}{*}{E} & \textit{FourPeople} & 0.012 & 0.265 & 0.010 & 0.231 & 0.010 & 0.234 & 0.010 & 0.238 & 0.008 & 0.219 & \textbf{-0.040} & \textbf{-1.129}\\
            \cline{3-15} & & \textit{Johnny} & 0.010 & 0.308 & 0.010 & 0.339 & 0.013 & 0.215 & 0.012 & 0.363 & 0.010 & 0.259 & \textbf{-0.065} & \textbf{-1.171}\\
            \cline{3-15} & & \textit{KristenAndSara} & 0.016 & 0.278 & 0.015 & 0.316 & 0.016 & 0.360 & 0.015 & 0.338 & 0.014 & 0.352 & \textbf{-0.026} & \textbf{-1.305}\\
            \cline{2-15} & \multicolumn{2}{c|}{Average} & 0.022 & 0.461 & 0.020 & 0.483 & 0.019 & 0.419 & 0.020 & 0.457 & 0.020 & 0.495 & \textbf{-0.049} & \textbf{-0.764}\\
            \hline
            \hline 37 & \multicolumn{2}{c|}{Average} & 0.027 & 0.541 & 0.026 & 0.621 & 0.024 & 0.562 & 0.027 & 0.617 & 0.024 & 0.614 & \textbf{-0.046} & \textbf{-0.993}\\
            \hline
        \end{tabular}
}
    \end{center}
\end{table}

In this section, we evaluate the performance of our method in terms of learned perceptual image patch similarity (LPIPS) \cite{zhang2018unreasonable} and perceptual index (PI) \cite{blau20182018}, which are metrics widely used for perceptual quality evaluation.
For results in terms of other evaluation metrics, please see the supplementary material.
We compare our method with Li \textit{et al.} \cite{li2017efficient}, DS-CNN \cite{yang2017decoder}, DCAD \cite{wang2017novel}, MFQE \cite{yang2018multi} and MFQE 2.0 \cite{guan2019mfqe}.
Among them, Li \textit{et al.} is the latest quality enhancement methods for compressed images.
DCAD and DS-CNN are single-frame enhancement methods for compressed videos, while MFQE and MFQE 2.0 are state-of-the-art multi-frame quality enhancement methods for compressed videos.
All these methods are trained over the same training set as ours for fair comparison.

Table \ref{metric_table} reports the $\Delta \rm {LPIPS}$ and $\Delta \rm {PI}$ results which are calculated between enhanced and compressed frames averaged over each test sequence.
Note that $\Delta \rm {LPIPS} <0$ and $\Delta \rm {PI} <0$ indicate improvement in perceptual quality.
As shown in this table, our MW-GAN outperforms all 6 compared methods in terms of perceptual quality, for all test sequences, while all the compared methods fail to enhance the perceptual quality of compressed video.
Specifically, at QP$=32$, the average $\Delta \rm {LPIPS}$ and $\Delta \rm {PI}$ in our MW-GAN are $-0.049$ and $-0.764$, respectively, while other methods \cite{zhang2017beyond,li2017efficient,yang2017decoder,wang2017novel,yang2018multi,guan2019mfqe} all have positive $\Delta \rm{LPIPS}$ and $\Delta \rm{PI}$ values.
Furthermore, our MW-GAN also gains the largest decrease of $\Delta \rm {LPIPS}$ with $-0.046$ and $\Delta \rm {PI}$ with $-0.993$ at QP$=37$, verifying the generalization ability of our MW-GAN for perceptual quality enhancement of compressed video.

\begin{figure}
    \begin{center}
        \centerline{\includegraphics[width=1.0\columnwidth]{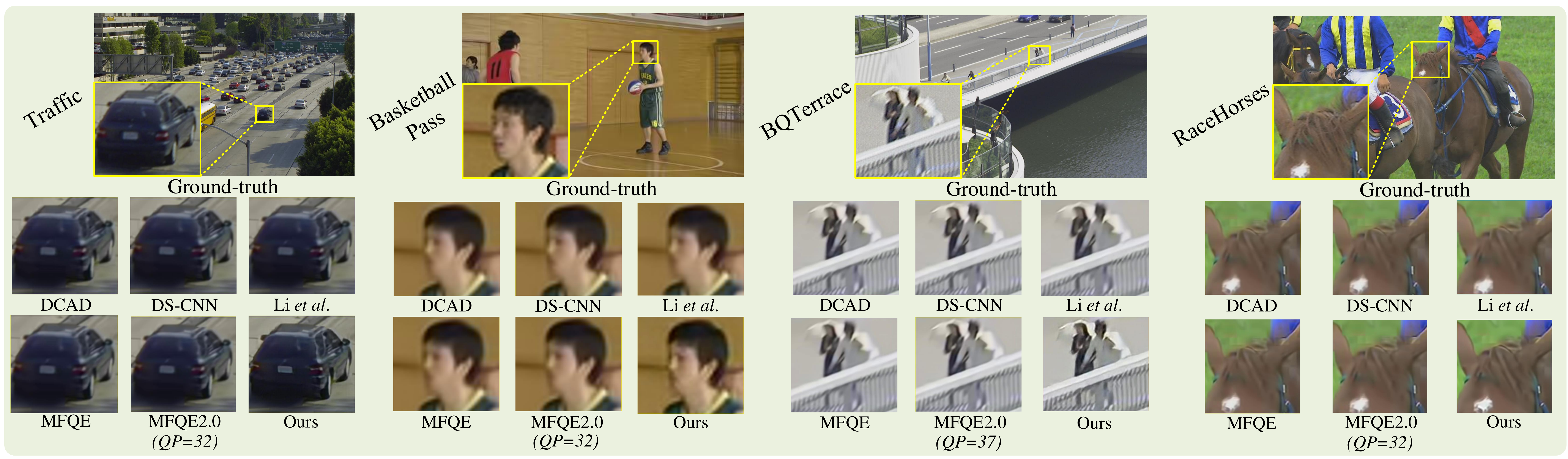}}
        \caption{Qualitative comparison on the test sequences of JCT-VC \cite{ohm2012comparison}. Our MW-GAN can generate much more realistic results (Zoom in for best view).}
        \label{visual_result}
    \end{center}
\end{figure}

\subsection{Subjective Comparison}
\label{exp_mos}
In this section, we mainly focus on the subjective evaluation of our method.
Figure \ref{visual_result} visualizes the enhanced video frames of \textit{Traffic} at QP$=32$, \textit{BasketballPass} at QP$=32$, \textit{BQTerrace} at QP $=37$ and \textit{RaceHorses} at QP$=32$.
We can observe that our MW-GAN method outperforms other methods with sharper edges and more vivid details.
For instance, the car in \textit{Traffic}, the face in \textit{BasketballPass}, the pedestrians in \textit{BQTerrace} and the horse in \textit{RaceHorses} can be finely restored in our MW-GAN, while the existing PSNR-oriented methods generate blurry results with low perceptual quality.

To further evaluate the subjective quality of our method, we also conduct a mean opinion score (MOS) test at QP$=37$.
Specifically, we asked 15 subjects to rate an integral score from $1$ to $100$ following \cite{seshadrinathan2010study} on the enhanced videos. The higher score indicates better perceptual quality.
The subjects are required to rate the scores for compressed video sequences, raw sequences, sequences enhanced by Li \textit{et al.} \cite{li2017efficient}, MFQE \cite{yang2018multi}, MFQE 2.0 \cite{guan2019mfqe}, our MW-GAN method.
Table \ref{mos_result} presents $\Delta \rm{MOS}$ which is calculated between enhanced and compressed sequences.
As can be seen, our MW-GAN method obtains the highest $\Delta \rm{MOS}$ score for each class of sequences, and it obtains an average increase of $4.55$ in terms of MOS, considerably better than other methods.

\begin{table}
    \newcommand{\tabincell}[2]{\begin{tabular}{@{}#1@{}}#2\end{tabular}}
    \begin{center}
        \caption{$\Delta \rm {MOS}$ calculated between enhanced and compressed sequences at QP $=37$ over the test sequences of JCT-VC \cite{ohm2012comparison}. Our MW-GAN obtains the highest $\Delta \rm{MOS}$ score for each class of sequences.}
        \label{mos_result}
    \resizebox{0.80\textwidth}{!}{
        \begin{tabular}{c|cccc}
            \hline  Sequence & Li \textit{et al.} \cite{li2017efficient} & MFQE \cite{yang2018multi} & MFQE 2.0 \cite{guan2019mfqe} & Ours \\
            \hline  A & -0.47 & -0.83 & -1.21 & \textbf{4.74} \\
            \hline  B & 1.86 & 1.42 & 0.82 & \textbf{3.95} \\
            \hline  C & -4.76 & -2.74 & -5.07 & \textbf{4.05} \\
            \hline  D & -1.08 & -0.30 & -1.90 & \textbf{5.15} \\
            \hline  E &  1.11 & 1.04 & 0.70 & \textbf{4.86} \\
            \hline  Average & -0.69 & -0.28 & -1.33 & \textbf{4.55} \\
            \hline
        \end{tabular}
    }
    \end{center}
\end{table}

\subsection{Ablation study}
\label{exp_abla}
In this section, we mainly analyze the effectiveness of WPT applied in each component of our MW-GAN.

\begin{figure}
    \begin{center}
        \centerline{\includegraphics[width=0.9\columnwidth]{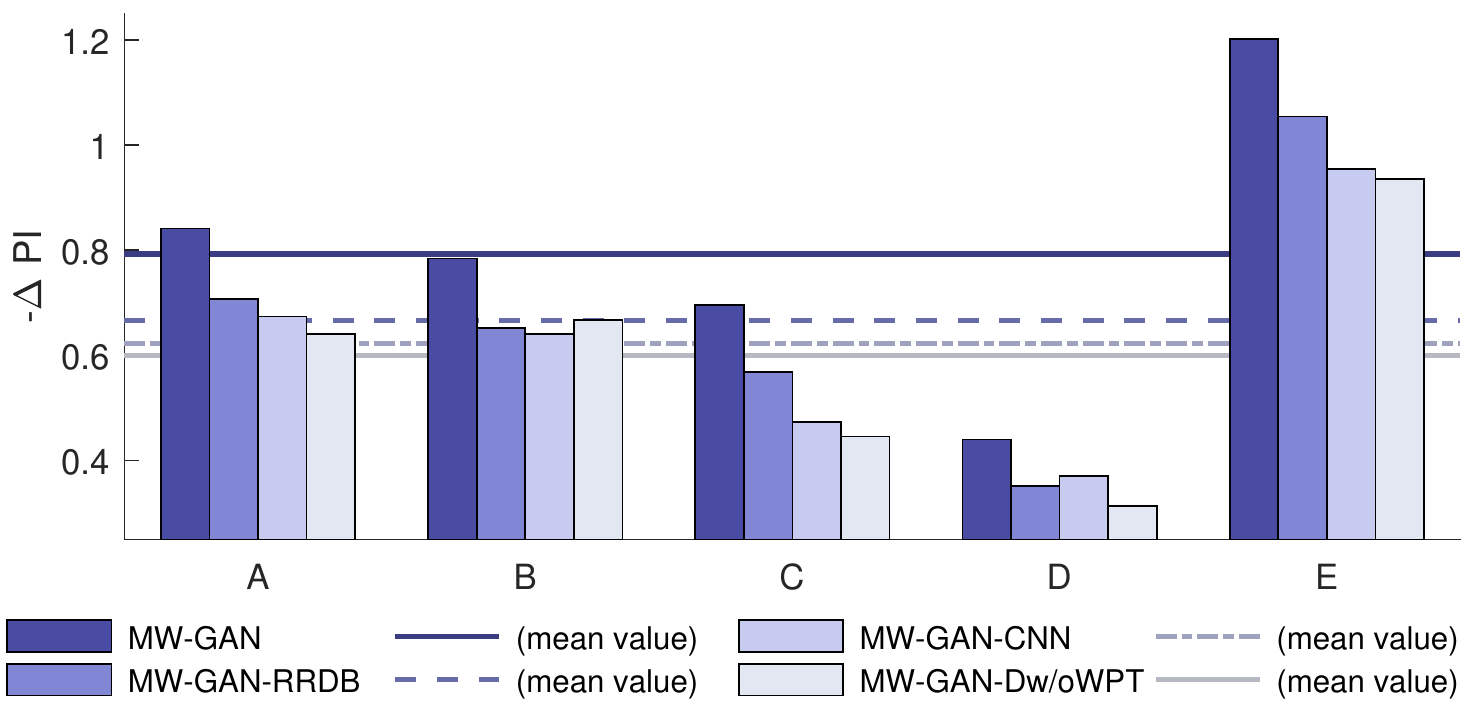}}
        \caption{Comparison between our MW-GAN and corresponding baselines on the test sequences of JCT-VC \cite{ohm2012comparison} according to $-\Delta \rm {PI}$ between enhanced and compressed sequences. Effectiveness of the proposed MW-GAN can be verified.}
        \label{abla_result}
    \end{center}
\end{figure}

\textbf{Effectiveness of WDRB in generator:}
In our generator, the multi-level WPT is achieved via adopting WDRB.
Thus, it is essential to validate the effectiveness of the proposed WDRB.
Here, we conduct the test with two strategies. (1) MW-GAN-RRDB: We directly replace the WDRB by the residual-in-residual dense block (RRDB) proposed in \cite{wang2018esrgan}. Note that the channels of the first convolutional layer before the first RRDB are increased to keep the structure of RRDB unchanged. (2) MW-GAN-CNN: We directly replace the WPT and IWPT in WDRB by average pooling and upsampling layers.
Note that the above two strategies all lead to parameter increasing.
Figure \ref{abla_result} shows the results of the above two ablation strategies over different classes of test sequences at QP$=32$.
We can see from this figure that WDRB has a positive impact on the perceptual quality, leading to $0.170$ and $0.126$ decrease in $\Delta \rm{PI}$ compared with MW-GAN-CNN and MW-GAN-RRDB, respectively.
Therefore, we can conclude that WDRB plays an effective role in MW-GAN.

\textbf{Effectiveness of WPT in the discriminator:}
To evaluate the effectiveness of the multi-level wavelet-based discriminator, we add a baseline without WPT in the discriminator, i.e., MW-GAN-D w/o WPT.
Specifically, we replace WPT in the discriminator with an average pooling layer that extracts features with the same size as wavelet sub-bands and keeps other components unchanged for a fair comparison.
As Figure \ref{abla_result} shows, without WPT in the discriminator, the $\Delta \rm{PI}$ score increases by average $0.192$ over the whole test set, which indicates the effectiveness of the proposed multi-level wavelet-based discriminator.

\section{Generalization ability}

In this section, we mainly focus on the generalization ability of the proposed
MW-GAN. For more details, please see our supplementary material.

\textbf{Transfer to H.264.}
To further verify the generalization capability of our MW-GAN, we also test on the video sequences compressed by other standards.
Specifically, we compress the test sequences of JCT-VC with H.264 (the JM encoder with the low-delay P configuration) at QP$=32$ and QP$=37$.
Then, we directly test the performance of our method over the above test sequences without fine-tuning.
Consequently, the average PI decrease of test sequences is 0.548 at QP$=32$ and 0.527 at QP=37, which implying the generalization ability of our MW-GAN approach across different compression standards.

\textbf{Performance on other sequences.}
In addition to the common-used test sequences of JCT-VC, we also test the performance of our and other methods over the test set in \cite{yang2018multi}, which is different from the test sequences of JCT-VC.
Results show that our method has 0.039 LPIPS decrease and 1.058 PI decrease at QP$=32$, while other methods all increase the PI value, which again indicates the superiority of our method according to the perceptual quality enhancement.

\textbf{Perception-distortion tradeoff.}
Our work mainly focuses on enhancing perceptual quality of compressed video.
However, according to the perception-distortion tradeoff \cite{blau2018perception}, improving perceptual quality fully will inevitably lead to a decrease of PSNR.
We further evaluate the objective quality of the frames generated by our method over the test sequences of JCT-VC.
Results show that our method leads to 1.040 dB PSNR decrease at QP$=32$ and 0.651 dB PSNR decrease at QP$=37$.
It is promising future work for perception-distortion tradeoff of compressed video quality enhancement, while at the current stage we mainly focus on perceptual quality enhancement (w/o considering PSNR) as the first attempt in this direction.

\section{Conclusion}
In this paper, we have proposed MW-GAN as the first attempt for perceptual quality enhancement of compressed video, which embeds WPT in both generator and discriminator for recovering high-frequency details.
First, we find from the data analysis that perceptual quality is highly related to the high-frequency sub-bands in wavelet domain.
Second, we design a wavelet-based GAN to recover high-frequency details in wavelet domain.
Specifically, a WDRB is proposed in our generator for wavelet sub-band reconstruction and a multi-level wavelet-based discriminator is applied to further encourage high-frequency recovery.
Finally, the ablation experiments showed the effectiveness of each component of MW-GAN.
More importantly, both quantitative and qualitative results in extensive experiments verified that our MW-GAN outperforms other state-of-the-art methods according to perceptual quality enhancement of compressed video.

\textbf{Acknowledgement}
This work was supported by the NSFC under Project 61876013, Project 61922009, and Project 61573037.

\clearpage
%
%
\bibliographystyle{splncs04}
\bibliography{egbib}

\begin{thebibliography}{10}
\providecommand{\url}[1]{\texttt{#1}}
\providecommand{\urlprefix}{URL }
\providecommand{\doi}[1]{https://doi.org/#1}

\bibitem{bampis2017study}
Bampis, C.G., Li, Z., Moorthy, A.K., Katsavounidis, I., Aaron, A., Bovik, A.C.:
  Study of temporal effects on subjective video quality of experience. IEEE
  Transactions on Image Processing (TIP)  \textbf{26}(11),  5217--5231 (2017)

\bibitem{blau20182018}
Blau, Y., Mechrez, R., Timofte, R., Michaeli, T., Zelnik-Manor, L.: The 2018
  pirm challenge on perceptual image super-resolution. In: Proceedings of the
  European Conference on Computer Vision (ECCV). pp.~0--0 (2018)

\bibitem{blau2018perception}
Blau, Y., Michaeli, T.: The perception-distortion tradeoff. In: Proceedings of
  the IEEE Conference on Computer Vision and Pattern Recognition (CVPR). pp.
  6228--6237 (2018)

\bibitem{cavigelli2017cas}
Cavigelli, L., Hager, P., Benini, L.: Cas-cnn: A deep convolutional neural
  network for image compression artifact suppression. In: 2017 International
  Joint Conference on Neural Networks (IJCNN). pp. 752--759. IEEE (2017)

\bibitem{chang2013reducing}
Chang, H., Ng, M.K., Zeng, T.: Reducing artifacts in jpeg decompression via a
  learned dictionary. IEEE Transactions on Signal Processing  \textbf{62}(3),
  718--728 (2013)

\bibitem{chu2018temporally}
Chu, M., Xie, Y., Leal-Taix{\'e}, L., Thuerey, N.: Temporally coherent gans for
  video super-resolution (tecogan). arXiv preprint arXiv:1811.09393  (2018)

\bibitem{cvni2019report}
CVNI: Cisco visual networking index: Global mobile data traffic forecast
  update, 2016-2021 white paper. In:
  https://www.cisco.com/c/en/us/solutions/collateral/service-provider/visual-networking-index-vni/white-paper-c11-741490.html
  (2017)

\bibitem{deng2019wavelet}
Deng, X., Yang, R., Xu, M., Dragotti, P.L.: Wavelet domain style transfer for
  an effective perception-distortion tradeoff in single image super-resolution.
  In: Proceedings of the IEEE International Conference on Computer Vision
  (CVPR). pp. 3076--3085 (2019)

\bibitem{dong2015compression}
Dong, C., Deng, Y., Change~Loy, C., Tang, X.: Compression artifacts reduction
  by a deep convolutional network. In: Proceedings of the IEEE International
  Conference on Computer Vision (ICCV). pp. 576--584 (2015)

\bibitem{dosovitskiy2016generating}
Dosovitskiy, A., Brox, T.: Generating images with perceptual similarity metrics
  based on deep networks. In: Advances in Neural Information Processing Systems
  (NIPS). pp. 658--666 (2016)

\bibitem{foi2007pointwise}
Foi, A., Katkovnik, V., Egiazarian, K.: Pointwise shape-adaptive dct for
  high-quality denoising and deblocking of grayscale and color images. IEEE
  Transactions on Image Processing (TIP)  \textbf{16}(5),  1395--1411 (2007)

\bibitem{guan2019mfqe}
Guan, Z., Xing, Q., Xu, M., Yang, R., Liu, T., Wang, Z.: Mfqe 2.0: A new
  approach for multi-frame quality enhancement on compressed video. IEEE
  Transactions on Pattern Analysis and Machine Intelligence (TPAMI) pp.~1--1
  (2019)

\bibitem{guo2016building}
Guo, J., Chao, H.: Building dual-domain representations for compression
  artifacts reduction. In: Proceedings of the European Conference on Computer
  Vision (ECCV). pp. 628--644. Springer (2016)

\bibitem{jancsary2012loss}
Jancsary, J., Nowozin, S., Rother, C.: Loss-specific training of non-parametric
  image restoration models: A new state of the art. In: Proceedings of the
  European Conference on Computer Vision (ECCV). pp. 112--125. Springer (2012)

\bibitem{jawerth1994overview}
Jawerth, B., Sweldens, W.: An overview of wavelet based multiresolution
  analyses. SIAM review  \textbf{36}(3),  377--412 (1994)

\bibitem{johnson2016perceptual}
Johnson, J., Alahi, A., Fei-Fei, L.: Perceptual losses for real-time style
  transfer and super-resolution. In: Proceedings of the European conference on
  computer vision (ECCV). pp. 694--711. Springer (2016)

\bibitem{jung2012image}
Jung, C., Jiao, L., Qi, H., Sun, T.: Image deblocking via sparse
  representation. Signal Processing: Image Communication  \textbf{27}(6),
  663--677 (2012)

\bibitem{kingma2014adam}
Kingma, D.P., Ba, J.: Adam: A method for stochastic optimization. International
  Conference on Learning Representations (ICLR)  (2015)

\bibitem{lai2017deep}
Lai, W.S., Huang, J.B., Ahuja, N., Yang, M.H.: Deep laplacian pyramid networks
  for fast and accurate super-resolution. In: Proceedings of the IEEE
  Conference on Computer Vision and Pattern Recognition (CVPR). pp. 624--632
  (2017)

\bibitem{ledig2017photo}
Ledig, C., Theis, L., Husz{\'a}r, F., Caballero, J., Cunningham, A., Acosta,
  A., Aitken, A., Tejani, A., Totz, J., Wang, Z., et~al.: Photo-realistic
  single image super-resolution using a generative adversarial network. In:
  Proceedings of the IEEE Conference on Computer Vision and Pattern Recognition
  (CVPR). pp. 4681--4690 (2017)

\bibitem{li2017efficient}
Li, K., Bare, B., Yan, B.: An efficient deep convolutional neural networks
  model for compressed image deblocking. In: 2017 IEEE International Conference
  on Multimedia and Expo (ICME). pp. 1320--1325. IEEE (2017)

\bibitem{li2015weight}
Li, S., Xu, M., Deng, X., Wang, Z.: Weight-based r-$\lambda$ rate control for
  perceptual hevc coding on conversational videos. Signal Processing: Image
  Communication  \textbf{38},  127--140 (2015)

\bibitem{liew2004blocking}
Liew, A.C., Yan, H.: Blocking artifacts suppression in block-coded images using
  overcomplete wavelet representation. IEEE Transactions on Circuits and
  Systems for Video Technology (TCSVT)  \textbf{14}(4),  450--461 (2004)

\bibitem{mallat1999wavelet}
Mallat, S.: A wavelet tour of signal processing. Elsevier (1999)

\bibitem{mao2017least}
Mao, X., Li, Q., Xie, H., Lau, R.Y., Wang, Z., Paul~Smolley, S.: Least squares
  generative adversarial networks. In: Proceedings of the IEEE International
  Conference on Computer Vision (CVPR). pp. 2794--2802 (2017)

\bibitem{mathieu2015deep}
Mathieu, M., Couprie, C., LeCun, Y.: Deep multi-scale video prediction beyond
  mean square error. arXiv preprint arXiv:1511.05440  (2015)

\bibitem{mechrez2018maintaining}
Mechrez, R., Talmi, I., Shama, F., Zelnik-Manor, L.: Maintaining natural image
  statistics with the contextual loss. In: Asian Conference on Computer Vision
  (ACCV). pp. 427--443. Springer (2018)

\bibitem{meng2018mganet}
Meng, X., Deng, X., Zhu, S., Liu, S., Wang, C., Chen, C., Zeng, B.: Mganet: A
  robust model for quality enhancement of compressed video. arXiv preprint
  arXiv:1811.09150  (2018)

\bibitem{ohm2012comparison}
Ohm, J.R., Sullivan, G.J., Schwarz, H., Tan, T.K., Wiegand, T.: Comparison of
  the coding efficiency of video coding standards¡ªincluding high efficiency
  video coding (hevc). IEEE Transactions on Circuits and Systems for Video
  Technology (TCSVT)  \textbf{22}(12),  1669--1684 (2012)

\bibitem{sajjadi2017enhancenet}
Sajjadi, M.S., Scholkopf, B., Hirsch, M.: Enhancenet: Single image
  super-resolution through automated texture synthesis. In: Proceedings of the
  IEEE International Conference on Computer Vision (ICCV). pp. 4491--4500
  (2017)

\bibitem{seshadrinathan2010study}
Seshadrinathan, K., Soundararajan, R., Bovik, A.C., Cormack, L.K.: Study of
  subjective and objective quality assessment of video. IEEE Transactions on
  Image Processing (TIP)  \textbf{19}(6),  1427--1441 (2010)

\bibitem{sullivan2012overview}
Sullivan, G.J., Ohm, J.R., Han, W.J., Wiegand, T.: Overview of the high
  efficiency video coding (hevc) standard. IEEE Transactions on Circuits and
  Systems for Video Technology (TCSVT)  \textbf{22}(12),  1649--1668 (2012)

\bibitem{tai2017memnet}
Tai, Y., Yang, J., Liu, X., Xu, C.: Memnet: A persistent memory network for
  image restoration. In: Proceedings of the IEEE International Conference on
  Computer Vision (ICCV). pp. 4539--4547 (2017)

\bibitem{wang2017novel}
Wang, T., Chen, M., Chao, H.: A novel deep learning-based method of improving
  coding efficiency from the decoder-end for hevc. In: 2017 Data Compression
  Conference (DCC). pp. 410--419. IEEE (2017)

\bibitem{wang2018recovering}
Wang, X., Yu, K., Dong, C., Change~Loy, C.: Recovering realistic texture in
  image super-resolution by deep spatial feature transform. In: Proceedings of
  the IEEE Conference on Computer Vision and Pattern Recognition (CVPR). pp.
  606--615 (2018)

\bibitem{wang2018esrgan}
Wang, X., Yu, K., Wu, S., Gu, J., Liu, Y., Dong, C., Qiao, Y., Change~Loy, C.:
  Esrgan: Enhanced super-resolution generative adversarial networks. In:
  Proceedings of the European Conference on Computer Vision (ECCV). pp.~0--0
  (2018)

\bibitem{wang2016d3}
Wang, Z., Liu, D., Chang, S., Ling, Q., Yang, Y., Huang, T.S.: D3: Deep
  dual-domain based fast restoration of jpeg-compressed images. In: Proceedings
  of the IEEE Conference on Computer Vision and Pattern Recognition (CVPR). pp.
  2764--2772 (2016)

\bibitem{xingjian2015convolutional}
Xingjian, S., Chen, Z., Wang, H., Yeung, D.Y., Wong, W.K., Woo, W.c.:
  Convolutional lstm network: A machine learning approach for precipitation
  nowcasting. In: Advances in Neural Information Processing Systems (NIPS). pp.
  802--810 (2015)

\bibitem{yang2019quality}
Yang, R., Sun, X., Xu, M., Zeng, W.: Quality-gated convolutional lstm for
  enhancing compressed video. In: 2019 IEEE International Conference on
  Multimedia and Expo (ICME). pp. 532--537. IEEE (2019)

\bibitem{yang2018enhancing}
Yang, R., Xu, M., Liu, T., Wang, Z., Guan, Z.: Enhancing quality for hevc
  compressed videos. IEEE Transactions on Circuits and Systems for Video
  Technology (TCSVT)  (2018)

\bibitem{yang2017decoder}
Yang, R., Xu, M., Wang, Z.: Decoder-side hevc quality enhancement with scalable
  convolutional neural network. In: 2017 IEEE International Conference on
  Multimedia and Expo (ICME). pp. 817--822. IEEE (2017)

\bibitem{yang2018multi}
Yang, R., Xu, M., Wang, Z., Li, T.: Multi-frame quality enhancement for
  compressed video. In: Proceedings of the IEEE Conference on Computer Vision
  and Pattern Recognition (CVPR). pp. 6664--6673 (2018)

\bibitem{zhang2017beyond}
Zhang, K., Zuo, W., Chen, Y., Meng, D., Zhang, L.: Beyond a gaussian denoiser:
  Residual learning of deep cnn for image denoising. IEEE Transactions on Image
  Processing (TIP)  \textbf{26}(7),  3142--3155 (2017)

\bibitem{zhang2018unreasonable}
Zhang, R., Isola, P., Efros, A.A., Shechtman, E., Wang, O.: The unreasonable
  effectiveness of deep features as a perceptual metric. In: Proceedings of the
  IEEE Conference on Computer Vision and Pattern Recognition (CVPR). pp.
  586--595 (2018)

\end{thebibliography}
\end{document}